\documentstyle[12pt]{article}
\newcommand{\rb}{|\!r}
\newcommand{\ra}{ r\!|}
\newcommand{\rc}{\backslash\!\!\!r}
\newcommand{\rd}{r\!\!\!/}

\title{The complete solution to the constant quantum
Yang-Baxter equation in two dimensions }
\author{Jarmo Hietarinta\\
Department of Physics, University of Turku,20500 Turku,
FINLAND\\email: hietarin@utu.fi}
\date{}

\begin{document}

\maketitle

\begin{abstract}
We describe how the complete solution to the two-dimensional constant
quantum Yang-Baxter equation was found [1-3].
\end{abstract}

\section{Introduction}
In this talk we give a brief description of how the two-dimensional
constant quantum Yang-Baxter equation (YBE) was solved. The results
were announced in [1], a detailed account of the solution process will
be published elsewhere [2], see also [3].

First let us recall the spectral parameter dependent form of YBE:
\begin{equation}
R_{j_1j_2}^{k_1k_2}(u)R_{k_1j_3}^{l_1k_3}(u+v)R_{k_2k_3}^{l_2l_3}(v)=
R_{j_2j_3}^{k_2k_3}(v)R_{j_1k_3}^{k_1l_3}(u+v)R_{k_1k_2}^{l_1l_2}(u).
\end{equation}
Here summation over the repeated $k$ indices is understood.
The constant quantum YBE
\begin{equation}
R_{j_1j_2}^{k_1k_2}R_{k_1j_3}^{l_1k_3}R_{k_2k_3}^{l_2l_3}=
R_{j_2j_3}^{k_2k_3}R_{j_1k_3}^{k_1l_3}R_{k_1k_2}^{l_1l_2}
\end{equation}
is obtained from (1) with $u=v=0$ or $u=v=\pm\infty$.

The YBE appears in many physical contexts.  It was first derived in
the study of solvable vertex models in statistical mechanics as the
condition of commuting transfer matrices (Yang, Baxter), another
derivation follows from the factorization of the S-matrix in 1+1
dimensional Quantum Field Theory (Zamolodchikov).  The YBE is also
essential in Quantum Inverse Scattering Method for integrable systems
as developed by the Leningrad school.  The more abstract setting is in
terms of Sklyanin algebras, quantum groups and Hopf algebras
(Drinfeld). More recently the connection to braid groups and knot
theory has been studied, here the spectral parameter independent form
arises naturally.  For a nice annotated collection of basic papers,
see [4].

For applications we then need solutions of YBE.  Many solutions have
been found before, either by Lie-algebraic methods [5], or by using a
specific ansatz [6]. In general YBE has $N^6$ cubic equations for
$N^4$ unknowns, so even in the simplest, i.e.\ two-dimensional case, one
has 64 equations for 16 unknowns.  Such a set of equations is
certainly too complicated for a brute force approach. In order
to find the complete solution it is therefore necessary to
simplify the problem by all means possible (without sacrifying
generality). This is done using the inherent symmetries of the
system.

In writing out the two indiced object $R$ we use the usual matrix
notation.  In two dimensions we have four $2\times2$ blocks, the
second index pair denotes the location of the block:
\begin{equation}
R=\left(\begin{array}{cccc}
R_{11}^{11} & R_{11}^{21} & R_{11}^{12} & R_{11}^{22} \\
R_{21}^{11} & R_{21}^{21} & R_{21}^{12} & R_{21}^{22} \\
R_{12}^{11} & R_{12}^{21} & R_{12}^{12} & R_{12}^{22} \\
R_{22}^{11} & R_{22}^{21} & R_{22}^{12} & R_{22}^{22}
\end{array}\right)
=
\left(\begin{array}{cccc}
a & b & c & d \\
f & g & h & j \\
k & l & m & n \\
p & q & u & v
\end{array}\right).
\end{equation}

\section{Symmetries}
The quantum YBE is invariant under the following continuous group of
transformations
\begin{equation}
R\to \kappa(Q\otimes Q) R (Q\otimes Q)^{-1},
\end{equation}
where $Q$ is a nonsingular $N\times N$ matrix and $\kappa$ a nonzero
number. One representative is sufficient of all solutions related by
(4), and during the solution process we will choose that
representative which produces the simplest equations.  We will use up
this rotational freedom in a specific order, using the following
parametrization:
\begin{equation}
Q=\left(\begin{array}{cccc} A & 0 \\ 0 & 1/A\end{array}\right)
\left(\begin{array}{cccc} 1 & 0 \\ C & 1\end{array}\right)
\left(\begin{array}{cccc} 1 & B \\ 0 & 1\end{array}\right).
\end{equation}
In particular note the scalings with $\kappa$ and a diagonal $Q$.
Let us define the scaling weight of a matrix element by $w(R_{ij}^{kl})=
k+l-i-j$, then two nonzero elements of $R$ with different weights can be
scaled to one using $A$ and $\kappa$.

The YBE has also discrete symmetries related to index changes:
\begin{eqnarray}
R_{ij}^{kl}&\to& R_{kl}^{ij},\\
R_{ij}^{kl}&\to& R^{k+n,l+n}_{i+n,j+n},\mbox{ (indices mod $N$) }\\
R_{ij}^{kl}&\to& R_{ji}^{lk}.
\end{eqnarray}
In two dimensions and using the matrix notation (3), (6) corresponds
to a reflection across the diagonal (= the usual transposition, also
called the P reflection), (7) with $n=1$ and followed by (6)
corresponds to a reflection across the secondary diagonal (C
reflection), and finally (8) corresponds to a reflection among the two
central rows and among the two central columns (T reflection).

\section{Breakup into smaller parts}
Since the system under stydy is so large we must first split it by
hand into several smaller subproblems. [When we refer to
specific equations, we use the following numbering:
$E_{l_3+2l_2+4l_1+8j_3+16j_2+32j_1-62}=
R_{j_1j_2}^{k_1k_2}R_{k_1j_3}^{l_1k_3}R_{k_2k_3}^{l_2l_3}-
R_{j_2j_3}^{k_2k_3}R_{j_1k_3}^{k_1l_3}R_{k_1k_2}^{l_1l_2}$.
The equations are written out explicitly in [2,3].]

i) First we analyzed the equations by counting how many times each
variable appears in them and it turned out that the corner elements
$d$ and $p$ appeared most frequently. Thus it seemed to be good idea
to transform so that $d=0$. This can always be done: If $p=0$ we use
the P reflection, else we use the $B$ part in (5) to get
\begin{equation}
\begin{array}{rl}
d_{new}:=&B^4p+B^3(f+k-q-u)+B^2(a-g-h-l-m+v)\\
&\quad+B(-b-c+j+n)+d,
\end{array}
\end{equation}
and since $p\ne0$ we can always find a $B$ so that $d_{new}=0$.

ii) At this point we had $d=0$ and to keep it that way we take $B=0$ in
subsequent transformations. When looking at the equations we found that
one of them had a nice form
\begin{equation}
 E_{22}:=bc(f-k)+jn(q-u)=0,
\end{equation}
and thus the problem would factorize into three parts, if we could
transform so that $f-k=0$ or $q-u=0$. This was accomplished as follows:

If $f=k$ already there is nothing to do, if $f\ne k$ but $q=u$ use C
reflection to put $f=k$, and in both case take $C=0$ in $Q$ (5).  If
both $f\ne k$ and $q\ne u$ we have after transforming with the $C$
part
\begin{equation}
\begin{array}{rl}
(f-k)_{new}:&=C^2(j-n)+C(-g-h+l+m)+f-k=0,\\
(q-u)_{new}:&=C^2(b-c)+C(g-h+l-m)+q-u=0.
\end{array}
\end{equation}
We can now solve for $C$ in one of the these (and use reflection (7)
if necessary), except if $j=n,\,b=c,\,h=l,\,g=m,\,f\ne k,\,q\ne u$,
which will become case C.

iii) When we use $d=0,\,f=k$ in $E_{22}$ we find that the problem
splits into two big cases A: $q=u$ and B: $n=0,\,q\ne u$.  (The case
$j=0$ can be T reflected (9) into $n=0$.) When these assignments are
used we find that some other equations factorize and the problem
splits into six simpler cases (A1, A2, B1, B2, B3, C).

v) The two remaining continuous freedoms are related to scalings by
$A$ and $\kappa$ (4,5). If there are two nonzero elements with
different weights they can both be scaled to unity. If we have
elements of equal weight only one of them can still be scaled to
unity. With the scalings we can split the problem further into a total
of 33 subcases.

\section{Computer solution}
The best way to analyze sets of polynomial equations is by using
Gr\"obner bases.  This is a systematic approach (Buchberger algorithm)
to sets of equations and defines a canonical form in a given ordering
of variables. It has been implemented in computer algebra systems.

Since we just need solutions to a set of equations it is a good
strategy to factorize the polynomials when possible and split the
problem into smaller ones. Thus for each of the 33 subsubcases we
computed the factorized Gr\"obner basis using the `groebner'-package
written by Melenk, M\"oller and Neun [7] for the REDUCE 3.4 [8].  The
raw output contained repeats and subcases which were eliminated by a
separate program. (In the newest version they are eliminated
automatically). In the end we had 96 solutions to analyze.

\section{Canonical form}
Since many of the solutions obtained above can be transformed into
each other it is important to bring them into a canonical form (using
the continuous and discrete symmetries discussed before) for a final
comparison.

In order to define a reasonable canonical form let us consider the
trace matrices of $R$:
\begin{equation}
\ra_i^k= R_{ij}^{kj},\quad
\rb_j^l=R_{ij}^{il},\quad
\rc_i^l= R_{ij}^{jl},\quad
\rd_j^k= R_{ij}^{ki}.
\end{equation}
Under (4) all of these transform according to
\begin{equation}
r\to \kappa Q r Q^{-1}.
\end{equation}
The basic definition of the proposed canonical form is that the above
trace matrices are in the Jordan canonical form. Since the trace
matrices do not necessarily commute it is possible that they cannot be
brought to the canonical form simultaneously, we will therefore work
in the above order. If this requirement is not enough to fix the
rotational freedom completely, we must look at individual matrix
elements of $R$.  (For a detailed algorithm for constructing the
canonical form, see [1,2].)  Using these ideas we were able to combine
the 96 solutions into 23 cases using homogeneous parametrization [1].

\section{Conclusions}
With the work [1,2] we know all solutions of constant quantum YBE in 2
dimensions.  Most of them fit into the 8-vertex ansatz, but not all.
One interesting observation is that all nonsingular solutions are
either upper triangular, or have the property $R_{ij}^{kl}=0$ unless
$i+j=k+l\pmod{2}$.

There are now many things that we can do with the results. For example,
what kind of algebras do we get with the new solutions? And what kind of
applications are related to them?

As for extensions, next one could search for solutions with a spectral
parameter. Preliminary studies indicate that this can also be done
systematically and completely, but the work involved seems to be
rather extensive.  Another interesting problem is to go to higher
dimensions.

\end{document}